\documentclass[journal=jpcafh,manuscript=article
,layout=twocolumn
]{achemso}

\makeatletter
\let\l@addto@macro\relax
\makeatother
\usepackage[fontsize=10pt]{scrextend}
\usepackage{chemformula} % Formula subscripts using \ch{}
\usepackage[T1]{fontenc} % Use modern font encodings
\usepackage{amsmath}
\usepackage{appendix}
\usepackage{booktabs}

\author{Richard Einsele}
\author{Xincheng Miao}
\author{Luca Nils Philipp}
\author{Roland Mitri\'c}
\email{roland.mitric@uni-wuerzburg.de}
\affiliation{Institut f\"ur Physikalische und Theoretische Chemie, Julius-Maximilians-Universit\"at,  Emil-Fischer Str. 42, 97074, W\"urzburg, Germany}

\title
  {DIALECT, a software package for exciton spectra and dynamics in large molecular assemblies from weak to strong light-matter coupling regimes}


\let\oldmaketitle\maketitle
\let\maketitle\relax
\SectionNumbersOn
\begin{document}

\twocolumn[
\begin{@twocolumnfalse}
\oldmaketitle
%%%%%%%%%%%%%%%%%%%%%%%%%%%%%%%%%%%%%%%%%%%%%%%%%%%%%%%%%%%%%%%%%%%%%
%% The abstract environment will automatically gobble the contents
%% if an abstract is not used by the target journal.
%%%%%%%%%%%%%%%%%%%%%%%%%%%%%%%%%%%%%%%%%%%%%%%%%%%%%%%%%%%%%%%%%%%%%
\begin{abstract}
The software package DIALECT is introduced, which provides the capability of calculating excited-state properties and nonadiabatic dynamics of large molecular systems and can be applied to simulate energy and charge-transfer processes in molecular materials. To this end, we employ the FMO-LC-TDDFTB methodology, which combines the use of the fragment molecular orbital approach with the density-functional tight-binding method and an excitonic Hamiltonian including local and charge-transfer excitations. 
In this work, we present the features and capabilities of the DIALECT software package in simulating the excited state dynamics of molecules and molecular aggregates using exemplary trajectory surface hopping as well as decoherence corrected Ehrenfest dynamics calculations in the framework of LC-TDDFTB and FMO-LC-TDDFTB. In addition, the capability of simulating the polaritonic excited state properties is highlighted by the calculation of the polariton dispersion of an aggregate of naphthalene molecules. The development of the DIALECT program will facilitate the investigation of exciton and charge transport in large and complex molecular systems, such as biological aggregates, nanomaterials and other complex organic molecular systems.
\end{abstract}
\end{@twocolumnfalse}
]

%%%%%%%%%%%%%%%%%%%%%%%%%%%%%%%%%%%%%%%%%%%%%%%%%%%%%%%%%%%%%%%%%%%%%
%% Start the main part of the manuscript here.
%%%%%%%%%%%%%%%%%%%%%%%%%%%%%%%%%%%%%%%%%%%%%%%%%%%%%%%%%%%%%%%%%%%%%
\section{Introduction}
The development and advancement of efficient quantum mechanical methodologies is necessary to facilitate the theoretical simulation of the interactions within large molecular systems in the field of biomolecular and materials sciences. Especially, the calculation of the excited-state properties and the simulation of molecular dynamics in the excited-state manifold requires fast and efficient methods\cite{frauenheim_atomistic_2002,dreuw_single-reference_2005,gonzalez_progress_2012,knepp_excited-state_2025}. 

The density-functional theory (DFT) is the usual choice for the theoretical simulation of medium sized molecular systems (up to a few hundred atoms), due to its broad applicability and relatively short calculation times. In addition, the linear response formalism of the time-dependent density-functional theory (TD-DFT) provides an efficient approach to simulate the excited-state properties of various molecular systems\cite{casida_time-dependent_1995,casida_progress_2012}. Nevertheless, the theoretical description of molecular systems with the TD-DFT method is still limited by the computational requirements of the method, reducing its applicability to molecular systems consisting of around a few hundred atoms.

Accordingly, various approximations to the TD-DFT method have been developed to increase the computational efficiency of the approach, among them the sTDA/sTD-DFT method\cite{grimme_simplified_2013,bannwarth_simplified_2014}, TD-DFT+TB\cite{ruger_tight-binding_2016}, linear-scaling TD-DFT\cite{zuehlsdorff_linear-scaling_2013,zuehlsdorff_linear-scaling_2015} and subsystem TD-DFT\cite{jacob_subsystem_2014,jacob_subsystem_2024}, which are implemented in software packages like TURBOMOLE\cite{franzke_turbomole_2023}, serenity\cite{niemeyer_subsystem_2023} and std2\cite{noauthor_std2_2025}, among others.  

The development of semiempirical quantum mechanical (SQM) methods, which are based on wavefunction or density-functional approaches, achieved the decrease in computational resources by neglecting the differential overlap between basis functions and utilizing minimal basis sets\cite{thiel_semiempirical_2014,christensen_semiempirical_2016}. While the first SQM approaches have been published in the 1970s and 80s\cite{dewar_ground_1977,dewar_development_1985,stewart_optimization_1989}, the development of the density-functional tight-binding (DFTB) method in the 1990s\cite{porezag_construction_1995,elstner_self-consistent-charge_1998} led to continuous developments in the domain of SQM techniques in the last 20 years. After Elstner et al. published the self-consistent charge (SCC) DFTB methodology\cite{elstner_self-consistent-charge_1998}, it has been extended to time-dependent DFTB (TD-DFTB) by Niehaus et al\cite{niehaus_tight-binding_2001}. Furthermore, the theoretical description of the charge interactions has been improved in the DFTB3 method\cite{gaus_dftb3_2011}, followed by the inclusion of the long-range correction to the DFTB method\cite{humeniuk_long-range_2015}, which enhanced the calculation of excited-state properties involving charge-transfer (CT) states. In addition, the method has been extended to nonadiabatic molecular dynamics in the excited-state manifold to simulate photochemical processes\cite{mitric_nonadiabatic_2009,humeniuk_dftbaby_2017,stojanovic_nonadiabatic_2017,inamori_spinflip_2020,uratani_fast_2020}. Various software packages based on the DFTB method have been developed, among them DFTB+\cite{hourahine_dftb_2020}, CP2K\cite{hutter_cp2k_2014}, DFTBaby\cite{humeniuk_dftbaby_2017}, ADF\cite{te_velde_chemistry_2001} and hotbit\cite{koskinen_density-functional_2009}. Complementary, the work of Grimme and coworkes on the extended tight-binding (xtb)\cite{grimme_robust_2017,bannwarth_gfn2-xtbaccurate_2019,bannwarth_extended_2021} methods has facilitated the accurate and efficient calculation of ground state properties of a multitude of molecular systems, owing to the parameterization of the methodology including over 80 different atoms. 

Despite the increased efficiency of the DFTB methodology, the approach is still limited to the simulation of molecular systems of several hundred up to a thousand atoms. Accordingly, fragment-based linear scaling methodologies like the fragment molecular orbital (FMO)\cite{kitaura_fragment_1999,fedorov_multilayer_2005,nakano_fragment_2000,nakano_fragment_2002} or the divide-and-conquer (DC)\cite{akama_implementation_2007,kobayashi_extension_2008,kobayashi_divide-and-conquer_2010} methods have been developed and implemented in various software packages like GAMESS-US\cite{zahariev_general_2023} and DC-DFTBMD\cite{nishimura_d_2019}. These techniques are based on the partitioning of a molecular system in several fragments, whose individual properties are combined with the interactions of the various fragments to approximate the properties of the complete system. While both theories have been combined with different quantum mechanical methods, the utilization of SQM techniques in conjunction with fragmentation methods has facilitated the simulation of molecular systems up to a million atoms\cite{nishimura_quantum_2021}. The application of the fragment molecular orbital DFTB (FMO-DFTB)\cite{nishimoto_density-functional_2014} and the divide-and-conquer DFTB (DC-DFTB)\cite{nakai_divide-and-conquer-type_2016} method has enabled the study of complex large molecular systems, such as organic semiconductors\cite{nishimoto_density-functional_2014} or biomolecules\cite{komoto_development_2019}. 
Further extensions of either of both methods include the polarizable continuum model\cite{nishimoto_fragment_2016}, the long-range correction\cite{vuong_fragment_2019,komoto_large-scale_2020} and periodic boundary conditions\cite{nishimoto_fragment_2021}. 
However, while the FMO and DC have been applied to investigate photochemical processes, the excitations were confined to singular fragments of the whole molecular systems, with the remaining part being treated on the ground-state level\cite{uratani_trajectory_2021,nakata_analytic_2023}.

Thus, to address the complete theoretical description of the excited-state manifold in the framework of the FMO-DFTB method, we combined the long-range corrected FMO-DFTB (FMO-LC-DFTB) method with an excitonic Hamiltonian to obtain all singly excited singlet states of large molecular assemblies\cite{einsele_long-range_2023}. We employ a quasi-diabatic basis consisting of locally excited (LE) and CT states, which are obtained from LC-DFTB calculations of the monomer and pair fragments of the complete molecular system. After the Hamiltonian is constructed from the state energies and the couplings between the basis states, the diagonalization yields the excited-state spectrum of a complete molecular system.

Furthermore, we have recently extended our FMO-LC-TDDFTB methodology to the simulation of nonadiabatic excited-state molecular dynamics by utilizing the Ehrenfest approach with the quasi-diabatic basis and implementing the gradients of the basis states\cite{einsele_nonadiabatic_2024}. We could demonstrate that our method is applicable to simulating the exciton dynamics in aggregates of organic molecules and the charge-transfer processes in semiconductors.
In addition, we have recently developed a methodology to calculate the excited states of molecular aggregates that are influenced by strong light-matter coupling to microcavities\cite{einsele_fmo-lc-tddftb_2024}.

In this work, we introduce our software package DIALECT (DIAbatic Locally Excited and Charge Transfer states), which is open source and freely available on Github\cite{noauthor_dialect_2025}. It has been designed to enable the efficient simulation of excited-state properties of large molecular systems and provide insights into exciton and charge-transfer dynamics of various molecular materials, such as organic semiconductors or biomolecular aggregates. To facilitate these calculations, we implemented the DFTB and FMO-LC-TDDFTB methods in combination with molecular dynamics. 

The DIALECT software package is written in the Rust programming language\cite{perkel_why_2020}, which provides a code and memory safe environment in conjunction with the speed and efficiency of programming languages like C/C++ and Fortran. Although the adoption of Rust in the community of computational sciences is still sparse, we expect the language to gain popularity in the coming years.

This paper is structured as follows: in section \ref{sec:features}, we present an overview over the features of DIALECT and give a very brief summary of the theoretical methodology. In section \ref{sec:results}, we present exemplary calculations of nonadiabatic dynamics and polariton dispersion to highlight the capabilities of our software package. Section \ref{sec:timings} gives an overview of computational benchmarks of DIALECT and in section \ref{sec:conclusion}, we conclude this work and present a short outlook.
\newcommand\numberthis{\addtocounter{equation}{1}\tag{\theequation}}
\section{Feature Overview}\label{sec:features}
This section provides a brief summary of the most important theoretical methodologies which are implemented in the DIALECT software. After a short overview of the DFTB, the FMO-DFTB and FMO-LC-TDDFTB formalism, we present the nonadiabatic molecular dynamics formalism in the framework of surface hopping and the Ehrenfest method. Finally, we show our approach for the calculation of strong light-matter interactions of large molecular systems that are coupled to microcavities.

In this work, atomic units are used, accompanied by the following notation convention: uppercase letters $A$, $B$ and $C$ represent atoms, while molecular fragments are denoted by uppercase letters $I$ through $L$ without indices. Matrix elements are indicated by uppercase letters with indices, and matrices are represented by bold uppercase letters. Molecular orbital indices are denoted by lowercase letters, while Greek letters signify atomic orbital indices.
\subsection{DFTB Methodology}
The expression for the ground-state energy in the framework of DFTB is generally obtained from the expansion of the total energy functional of the electron density in DFT\cite{koskinen_density-functional_2009}. By expanding the DFT energy around a reference density, which is a superposition of the electron densities of the neutral atoms, up to the first\cite{porezag_construction_1995}, second\cite{elstner_self-consistent-charge_1998}, or third order\cite{gaus_dftb3_2011}, the working equations for the different approximations of tight-binding DFT are derived. Here, we only provide a very brief overview of the DFTB equations, for a more detailed derivation, we refer to other works\cite{elstner_self-consistent-charge_1998,koskinen_density-functional_2009,gaus_dftb3_2011,humeniuk_long-range_2015}. 

The expansion up to the first order yields the DFTB1 energy\cite{porezag_construction_1995}
\begin{equation}
\label{eq:dftb1}
E^{\text{DFTB1}} = \sum_{\mu\nu}P_{\mu\nu}H^0_{\mu\nu} + \sum_{AB}V^{rep}_{AB},
\end{equation}
where $P_{\mu\nu}$ is a matrix element of the electron density matrix, $H^0_{\mu\nu}$ a matrix element of the one-electron Hamiltonian and $V^{rep}_{AB}$ represents the repulsive energy contribution between pairs of atoms. $\mathbf{H^0}$ is calculated from precomputed and parameterised Slater-Koster tables by utilizing Slater-Koster transformations. This gives rise to a non-self-consistent method suitable only to qualitative studies.

The SCC DFTB\cite{elstner_self-consistent-charge_1998} (DFTB2) formalism includes density fluctuations up to second order, which are approximated by pairwise interactions of charge monopoles, resulting in the following energy expression:
\begin{align*}
\label{eq:dftb2}
E^{\text{DFTB2}} = &\sum_{\mu\nu}P_{\mu\nu}H^0_{\mu\nu} + \frac{1}{2}\sum_{AB}\gamma_{AB}\Delta q_{A}\Delta q_{B} \\ &+ \sum_{AB}V^{rep}_{AB}
\end{align*}
The interaction between two s-type charge densities on $A$ and $B$ is represented by the matrix element $\gamma_{AB}$, which is commonly derived from interactions between two Slater or Gaussian functions. If one considers Gaussian functions, $\gamma_{AB}$ is defined as 
\begin{equation}
\gamma_{AB} = \frac{\text{erf}(C_{AB}R_{AB})}{R_{AB}},
\end{equation}
where $R_{AB}$ is the distance between the atoms $A$ and $B$ and $C_{AB}= \left(2 (\sigma_A^2 + \sigma_B^2) \right)^{-\frac{1}{2}}$. $\sigma_A$ and $\sigma_B$ are the widths of the charge clouds on both atoms, which depend on the element-specific Hubbard parameters $U_A$ and $U_B$ as $\sigma_A=(\sqrt{\pi} U_A)^{-1}$. A detailed overview for the expression of $\gamma_{AB}$ using Slater functions is given by Gaus et al\cite{gaus_dftb3_2011}.
	
$\Delta q_A$ is the Mulliken charge difference on atom $A$ from the charge of the neutral reference atom and is defined as
\begin{equation}
	\Delta q_A = q_A - q^0_A= \sum_{\mu \in A ,\nu} \left[ P_{\mu\nu} S_{\mu\nu} - P^0_{\mu\nu} S_{\mu\nu}\right],
\end{equation}
where $\mathbf{S}$ is the overlap matrix, which is calculated from precomputed Slater-Koster tables analogously to $\mathbf{H^0}$.

Gaus et al. introduced the DFTB3 formalism\cite{gaus_dftb3_2011}, which includes charge fluctuations up to the third order to improve the DFTB methodology. The total energy expression is defined as 
\begin{align*}\label{eq:dftb3}
		E^{\text{DFTB3}} = &\sum_{\mu\nu}P_{\mu\nu}H^0_{\mu\nu} + \frac{1}{2}\sum_{AB}\gamma_{AB}\Delta q_{A}\Delta q_{B} \\ + &\frac{1}{3}\sum_{AB}\Gamma_{AB}\Delta q_{A}^2 \Delta q_{B} + \sum_{AB}V^{rep}_{AB},
\end{align*}
where $\Gamma_{AB}$ is the derivative of $\gamma_{AB}$ with respect to the Mulliken charge difference $\Delta q_A$.

A further improvement to the DFTB2 method was introduced by Humeniuk et al. by incorporating the long-range correction into the tight-binding DFT formalism\cite{humeniuk_long-range_2015,humeniuk_dftbaby_2017}. The Taylor expansion of a long-range corrected DFT functional up to the second order yields the following expression for the LC-DFTB2 energy:
\begin{align*}
		E^{\text{LC-DFTB2}} &= \sum_{\mu\nu}P_{\mu\nu}H^0_{\mu\nu} + \frac{1}{2}\sum_{AB}\gamma_{AB}\Delta q_{A}\Delta q_{B} \\ & -\frac{1}{16} \sum_{\mu \nu \lambda \sigma} \Delta P_{\mu \sigma} \Delta P_{\lambda \nu} S_{\mu \lambda} S_{\sigma \nu} \times \\ &\left[ \gamma_{\mu \sigma }^{lr} + \gamma_{\mu \nu}^{lr} + \gamma_{\lambda \sigma}^{lr} + \gamma_{\lambda \nu}^{lr} \right]  +\sum_{AB}V^{rep}_{AB}
\end{align*}\label{eq:lc-dftb2}
Here, $\mathbf{\Delta P}$ is the difference density matrix to the reference system of neutral atoms, and the long-range $\gamma$-matrix is defined as 
\begin{equation}
	\gamma^{lr}_{AB} = \frac{\text{erf}(C_{AB}^{lr})R_{AB}}{R_{AB}},
\end{equation}
where $\gamma_{\mu \nu} = \gamma_{AB}$ for $ \mu \in A$ and $\nu \in B$, and the long-range $C_{AB}^{lr}$ term
\begin{equation}
	C_{AB}^{lr} = \left[2 \left(\sigma_A^2 + \sigma_B^2 + \frac{1}{2} R_{lr}^2 \right)\right]^{-\frac{1}{2}}
\end{equation}
is dependent on the long-range radius $R_{lr}$.

Due to the fact that DFTB is a semiempirical method, its applicability is strongly influenced by the quality of the parameterization. In general, the parameterization process involves the generation of the electronic parameters, which are utilized to calculate $\mathbf{S}$ and $\mathbf{H^0}$, and the fitting of the repulsive potentials. We will not provide an outline of the parameterization procedure and the generation of the Slater-Koster files, as a detailed overview is available in Refs. \citenum{koskinen_density-functional_2009} and \citenum{vuong_parametrization_2018}. 

The DIALECT program supports the use of the DFTB2, DFTB3 and LC-DFTB methods, and both Slater and Gaussian functions for the calculation of the $\gamma$-matrix with atom-resolved as well as shell-resolved charges. 
\subsection{FMO-DFTB}
The FMO method employs a partitioning scheme of molecular systems into various fragments to increase the computational efficiency of quantum chemical approaches. It has been extensively documented in a number of papers\cite{kitaura_fragment_1999,fedorov_multilayer_2005,nakano_fragment_2000,nakano_fragment_2002}, and thus, we only provide a very short summary of the formalism. In case of the FMO-DFTB and FMO-LC-DFTB methods developed by Nishimoto et al.\cite{nishimoto_density-functional_2014} and Vuong et al\cite{vuong_fragment_2019}, the total ground-state energy is defined as 
\begin{equation}
	E = \sum_{I}^{N}E_I + \sum_{I,J>I} \left(E_{IJ} - E_{I} -E_{J} \right) + \sum_{I,J>I} \Delta E_{IJ}^{em},
\end{equation}
where $E_I$ is the (LC-)DFTB energy of a monomer fragment and $E_{IJ}$ the (LC-)DFTB energy of a dimer fragment of the complete system. The embedding energy difference between pair and monomer fragments is
\begin{equation}
	\Delta E_{IJ}^{em} = \sum_{A \in IJ} \sum_{K \neq I,J}\sum_{C \in K} \gamma_{AC} \Delta \Delta q_A^{IJ} \Delta q_C^{K}.
\end{equation}
The Mulliken charge difference between the dimer and its monomer fragments is $\Delta \Delta q_{A}^{IJ}$. 

A further reduction in computational resources in conjunction with only a slight decrease in accuracy is achieved by approximating the dimer energy for far-separated monomer fragments. This is called electrostatic-dimer (ES-DIM)\cite{nakano_fragment_2002} approximation and yields the following expression for the dimer energy:
\begin{equation}
	E_{IJ} = E_{I} + E_{J} + \sum_{A \in I}\sum_{B \in J} \gamma_{AB} \Delta q_A^I \Delta q_B^J
\end{equation}

Although the calculation of the total ground-state  energy is separated in monomer and dimer contributions, the SCC procedure of the FMO-DFTB method involves the computation of all monomer contributions collectively during each iteration, due to the fact that the monomer Hamiltonian (with LC)
\begin{align}
	H_{\mu \nu}^{I} &= H_{\mu \nu}^{0,I} + \frac{1}{2} S^I_{\mu \nu} \sum_{C} \left(\gamma^I_{AC}+\gamma^I_{BC} \right) \Delta q^I_C  + V_{\mu \nu }^I\\ &-\frac{1}{8} \sum_{\lambda \sigma} \Delta P_{\lambda \sigma} S_{\mu \lambda} S_{\sigma \nu} \left[ \gamma_{\mu \sigma}^{lr} + \gamma_{\mu \nu}^{lr} + \gamma_{\lambda \sigma}^{lr} + \gamma_{\lambda \nu}^{lr} \right]
\end{align}
contains the Coulomb interaction with all other monomers,
\begin{equation}
	V_{\mu \nu}^I = \frac{1}{2} S^I_{\mu \nu} \sum_{K \neq I}\sum_{C \in K} \left(\gamma^{IK}_{AC}+\gamma^{IK}_{BC} \right) \Delta q^K_C.
\end{equation}

A detailed overview of implementation of the simultaneous monomer SCC routines is presented in an earlier work\cite{einsele_long-range_2023}. As of now, the DIALECT program package only supports the FMO2 method, separating the system in monomer and dimer fragments, in conjunction with the DFTB2 and LC-DFTB2 formalism.

\subsection{FMO-LC-TDDFTB}
In order to extend the FMO-DFTB formalism to the calculation of electronically excited states of large molecular systems, we developed the FMO-LC-TDDFTB approach in a previous work\cite{einsele_long-range_2023}. In this subsection, we only provide a very brief summary of the methodology and refer to our previous publications\cite{einsele_long-range_2023,einsele_nonadiabatic_2024} for more details.

The wavefunction of a singly excited determinant in the framework of FMO-LC-TDDFTB is defined as the linear combination of quasi-diabatic locally excited singlet states on individual fragments and charge transfer states between fragment pairs:
\begin{equation}\label{eq:wavefunc_fmo}
	\left|\Psi \right\rangle = \sum_{I}^N \sum_{m}^{N_{LE}} c_{I}^{m} \left| \mathrm{LE}_{I}^{m}\right\rangle + \sum_{I}^{N} \sum_{J \neq I}^{N} \sum_{m}^{N_{CT}} c_{I \to J}^{m} \left| \mathrm{CT}_{I \to J}^{m} \right\rangle
\end{equation}
The quasi-diabatic LE and CT states are obtained from (LC)-TDA-DFTB calculations from monomer and pair fragments. The Hermitian eigenvalue problem 
\begin{equation}
	\mathbf{A}\mathbf{X} = \mathbf{\Omega}\mathbf{X}
\end{equation}
is solved in the (LC) tight-binding DFT formalism, where matrix elements of $\mathbf{A}$ are defined as 
\begin{equation}
	A_{ij,ab} = \delta_{ij} \delta_{ab} (\epsilon_a - \epsilon_i) + 2 \sum_{A,B} q_{A}^{ia} \gamma_{AB} q_{B}^{jb} - \sum_{A,B} q_{A}^{ij} \gamma_{AB}^{lr} q_{B}^{ab},
\end{equation}
and the atomic transition charges between MOs are
\begin{equation}
	q_{A}^{ij} = \frac{1}{2} \sum_{\mu \in A}\sum_{\nu} \left(C_{\mu i} C_{\nu j} + C_{\nu i} C_{\mu j} \right) S_{\mu \nu}.
\end{equation}

\begin{figure}[t!]
    \centering
    \includegraphics[width=\linewidth]{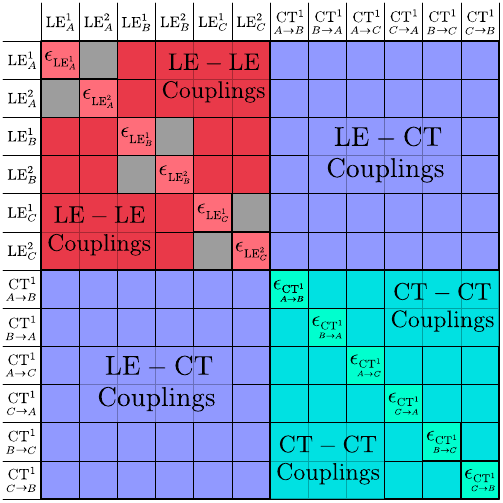}
    \caption{Excitonic Hamiltonian for an example system of three monomers with two locally excited states and one charge-transfer state for each fragment.}
    \label{fig:excitonic_hamiltonian}
\end{figure}

To obtain the expansion coefficients $c_{I}^{m}$ and $c_{I \to J}^{m}$ of the quasi-diabatic basis states, the complete excited-state Hamiltonian that is composed of all LE and CT states must be diagonalized. While the diagonal elements of the Hamiltonian are represented by the energies of the quasi-diabtic states, the off-diagonal elements require the calculation of the excitonic couplings between the basis states, yielding the excitonic Hamiltonian, which is depicted in Fig. \ref{fig:excitonic_hamiltonian} for an example system.

If only a part of the full manifold of excited states is requested, the Davidson algorithm is employed to obtain the singly excited singlet state energies\cite{davidson_iterative_1975}. Otherwise, a full diagonalization of the excitonic Hamiltonian is performed.
The couplings between the basis states are represented by the off-diagonal elements of the excitonic Hamiltonian. The off-diagonal matrix elements are split in one-electron and two-electron contributions. According to the Slater-Condon rules, the one-electron contribution vanishes for the LE-LE coupling, resulting in the following expression in the tight-binding formalism
\begin{align*}
\label{le_le_coupling}
\left\langle \mathrm{LE}_{I}^{m}\right|\mathrm{H}& \left| \mathrm{LE}_{J }^{n}\right\rangle 
= 2 \sum_{A\in I} \sum_{B \in J} q_{\mathrm{tr}, A}^{m(I)} \gamma_{AB} q_{\mathrm{tr}, B}^{n(J)} \\
-& \sum_{A\in IJ} \sum_{B \in IJ} 
\sum_{ia \in I} \sum_{jb \in J} X_{ia}^{m(I)} X_{jb}^{n(J)} q_{A}^{ij} \gamma_{AB}^{\mathrm{lr}} q_{B}^{ab}, \numberthis 
\end{align*}
where 
\begin{equation}
    q_{\mathrm{tr}, A}^{m(I)} = \sum_{ia} q_{A}^{ia} X_{ia}^{m(I)} 
\end{equation}
is the transition charge of the $m$-th excited state for the fragment $I$ on atom $A$. If the fragment pair of $I$ and $J$ is classified as a far-separated pair, the ES-DIM approximation is utilized and the second part of Eq. \ref{le_le_coupling} is neglected.

In contrast to the LE-LE interaction, the coupling between LE and CT basis states contains both one-electron and two-electron contributions. It is given by 
\begin{align*}
\label{le_ct_coupling}
\left\langle\mathrm{\mathrm{LE}}_{I}^{m}\right|\mathrm{H}& \left| \mathrm{CT}_{J \to K}^{n}\right\rangle = \\& \delta_{I J} \sum_{ia \in I}\sum_{b \in K} X_{ia}^{m(I)} X_{ib}^{n(I \to K)} H_{a b}^{\prime} \\ &-\delta_{I K}
    \sum_{ia \in I}\sum_{j \in J} X_{ia}^{m(I)}X_{ja}^{n(J \to I)} H_{i j}^{\prime} \\ 
+& 2\sum_{A \in I}\sum_{B \in JK} q_{\mathrm{tr}, A}^{m(I)} \gamma_{AB} q_{\mathrm{tr}, B}^{n(J \to K)} \\
- \sum_{A \in IJ}\sum_{B \in IK
} &\sum_{ia \in I}  \sum_{j \in J}\sum_{b \in K}X_{ia}^{m(I)} X_{jb}^{n(J \to K)} q_{A}^{ij} \gamma_{AB}^{\mathrm{lr}} q_{B}^{ab},  \numberthis 
\end{align*}
where $H^{'}_{ab}$ and $H^{'}_{ij}$ are matrix elements of the orthogonalized ground-state Hamiltonian of the complete system in the FMO method, which is defined as
\begin{equation}
    \mathbf{H}^{\prime} = \mathbf{S}^{-\frac{1}{2}} \mathbf{H}^{\mathrm{LCMO}} \mathbf{S}^{-\frac{1}{2}}.
    \label{eq:orthogonalized_hamiltonian}
\end{equation}
$\mathbf{S}$ is the overlap matrix of the complete system and 
\begin{equation}
    \mathbf{H}^{\mathrm{LCMO}} = \bigoplus_{I} \mathbf{H}^{I} + \bigoplus_{I>J}(\mathbf{H}^{IJ} - \mathbf{H}^{I} \oplus \mathbf{H}^{J})
\end{equation}
is the non-orthogonalized Hamiltonian that is built from the Hamiltonian matrices of monomer and pair fragments. The inverted overlap matrix is approximated in first order according to
\begin{equation}
    \mathbf{S}^{-\frac{1}{2}} \approx \frac{3}{2} \mathbf{1} - \frac{1}{2} \mathbf{S}.
\end{equation}
In case of far-separated fragments, the ES-DIM approximation is employed and the last part of Eq. \ref{le_ct_coupling} is neglected.

The coupling between two different charge-transfer basis states is given by
\begin{align}
\label{ct_ct_coupling}
\begin{split}
\left\langle \mathrm{CT}_{I \to J}^{m} \right|& \mathrm{H}\left| \mathrm{CT}_{K \to L}^{n}\right\rangle = \\ &2 \sum_{A\in IJ} \sum_{B \in KL} q_{\mathrm{tr}, A}^{m(I \to J)} \gamma_{AB} q_{\mathrm{tr}, B}^{n(K \to L)} \\
&- \sum_{i \in I}\sum_{a \in J}\sum_{j \in K}\sum_{b \in L} \sum_{A \in IK} \sum_{B \in JL} X_{ia}^{m(I \to J)}\\
&\times X_{jb}^{n(K \to L)}q_{A}^{ij} \gamma_{AB}^{\mathrm{lr}} q_{B}^{ab}. 
\end{split}
\end{align}
If fragment pairs between $I$ and $K$ or $J$ and $L$ are classified as ES-DIM pairs, the second term of the coupling is neglected and only the Coulomb term remains.

As of now, the DIALECT program only supports the use of the Tamm-Dancoff approximation for the calculation of the excited states of large molecular systems, as the full Casida linear-response formalism\cite{casida_time-dependent_1995} would require more computational resources due to its higher complexity.

\subsection{Surface Hopping Dynamics}\label{sec:surface_hopping}
In the present section, we briefly summarize the theoretical foundations of nonadiabatic molecular dynamics in the framework of surface hopping combined with DFTB as implemented in the DIALECT code. 

Surface hopping is a quantum-classical approach, where Newton's equations define the motion of the atomic nuclei, while the electronic dynamics is determined by quantum mechanics\cite{tully_molecular_1990,tully_nonadiabatic_1991}. For a fixed nuclear geometry, a quantum chemical calculation yields a manifold of electronic wavefunctions with their respective adiabatic energies $E_n$, which are parametrically dependent on the nuclear coordinates. The total electronic wavefunction can be expanded as a linear combination of the adiabatic states 
\begin{equation}
\label{eq:adiabatic_linear_comb}
    \left|\Psi(\mathbf{r};\mathbf{R}(t))\right\rangle = \sum_{n}c_n(t)\left|\psi_{n}(\mathbf{r};\mathbf{R}(t)) \right\rangle,
\end{equation}
where $c_n(t)$ are the time-dependent expansion coefficients of the adiabatic eigenstates. Inserting eq. \ref{eq:adiabatic_linear_comb} into the time-dependent Schr\"odinger equation yields a set of coupled differential equations for the expansion coefficients
\begin{equation}
    \text{i}\hbar\frac{\text{d}c_n}{\text{d}t} = E_n(\mathbf{R}(t) c_n(t)-\sum_m \text{i}\hbar D_{nm}(\mathbf{R}(t)) c_m(t),
\end{equation}
where $D_{nm}$ is the nonadiabatic coupling between two adiabtic states, which can be calculated from the nonadiabtic coupling vector and the nuclear velocity
\begin{equation}
    D_{nm}(\mathbf{R}(t)) = \left\langle \psi_n(\mathbf{r};\mathbf{R}(t)) \left| \nabla_R \right| \psi_m(\mathbf{r};\mathbf{R}(t)) \right\rangle \cdot \frac{\text{d}\mathbf{R}}{\text{d}t}.
\end{equation}
For the derivation of the nonadiabatic coupling vectors in the framework of the (LC-)TD-DFTB and (LC-)TDA-DFTB methodology, we refer to the references \citenum{einsele_nonadiabatic_2024} and \citenum{niehaus_exact_2023}.

Due to the stochastic nature of surface hopping, the simulation of nonadiabatic dynamics requires the calculation of an ensemble of nuclear trajectories, which is usually initialized from a phase-space distribution, e.g. a Wigner distribution\cite{bonacic-koutecky_theoretical_2005}. The nuclear trajectories are then propagated on a manifold of potential energy surfaces, which is calculated "on the fly" for the current nuclear geometry. The trajectories are restricted to move on a single adiabatic state, but a hop can transfer them to another state. The hopping probability from the current electronic state $i$ to the state $j$ is calculated according to\cite{lisinetskaya_simulation_2011}
\begin{equation}
    P_{i \to j} = \Theta(-\dot{\rho_{ii}})\Theta(\dot\rho_{jj})\frac{-\dot{\rho_{ii}}\dot\rho_{jj}}{\rho_{ii}\sum_{k}\Theta(\dot{\rho}_{kk})\dot{\rho}_{kk}}\Delta t,
\end{equation}
where $\Theta(x)$ is the Heaviside step function, $\mathbf{\rho}$ is the electronic density matrix
\begin{equation}
    \rho_{nm}(t) = c_n^*(t) c_m(t)
\end{equation}
and $\Delta t$ is the time step size of the nuclear propagation. In case of a hop between two adiabatic states, the potential energy has a discontinuity, and thus, the momentum is rescaled in the direction of the nonadiabatic coupling vector to preserve the energy conservation. If the hop occurs to a higher adiabatic state and the kinetic energy is lower than the potential energy difference, the hop is rejected.

A problem of the simulation of nonadiabatic dynamics in the framework of TD-DFTB is the lack of conical intersections between the ground state and excited states. Due to the fact that TD-DFT and TD-DFTB have zero coupling between the ground state and any excited state according to Brilluoin's theorem, the dimensionality of the intersection of the electronic states is reduced from the correct $N-2$ degrees of freedom to $N-1$, which results in the wrong shape of the state intersection. Our implementation tries to alleviate this error by a treating the hops to the ground state in a particular way: if the energy gap to the ground state is below a certain threshold, the transition to the ground state is forced. Additionally, after the ground state is reached, further hops to excited states are prohibited and the trajectory is solely propagated on the ground state potential energy surface.

In order to account for decoherence effects in surface hopping simulations, we have implemented the energy-based decoherence correction of Granucci et al.\cite{granucci_critical_2007,granucci_including_2010} in the DIALECT program. In this method, during the surface hopping simulation, the populations $\rho_{jj}$ of the adiabatic states ($j \neq i$) besides the currently occupied state ($i$) are damped by the factor
\begin{equation}
    e^{-\Delta t/\tau_{ji}},
\end{equation}
where 
\begin{equation}
    \tau_{ji} = \frac{\hbar}{|E_j - E_i|} \left(1 + \frac{C}{E_{\mathrm{kin}}} \right),
\end{equation}
$E_{\mathrm{kin}}$ is the kinetic energy and $C$ is a constant that is usually set to $0.1$~hartree. Subsequently, the population of the current adiabatic state $i$ is increased according to the differences in populations $\rho_{jj}$ before and after damping. 

%\textcolor{red}{Results of example system}
\subsection{Ehrenfest Exciton Dynamics}
We combined the Ehrenfest method with the FMO-LC-TDDFTB formalism in a previous work to facilitate the simulation of excitonic nonadiabatic molecular dynamics in large molecular systems\cite{einsele_nonadiabatic_2024}, for which the surface hopping approach is not feasible due to a large number of excited states leading to poor statistics. In this subsection, we give a short summary of the methodology and present the most important equations. %and present the simulation of the excited state dynamics of an anthracene chain.

In analogy to surface hopping, the Ehrenfest approach is a quantum-classical method, where the nuclear dynamics are treated classically, while the electronic dynamics is governed by quantum mechanics\cite{tully_mixed_1998}. In the case of the FMO-LC-TDDFTB methodology, the total electronic wavefunctions is expanded in the basis of the quasi-diabatic LE and CT states according to
\begin{equation}
\label{eq:diabatic_expansion}
        \left|\Psi(\mathbf{r};\mathbf{R}(t))\right\rangle = \sum_{n}c^{\mathrm{dia}}_n(t)\left|\psi_{n}^{dia}(\mathbf{r};\mathbf{R}(t)) \right\rangle,
\end{equation}
where $\psi^{\mathrm{dia}}$ are the quasi-diabatic basis states and $c^{\mathrm{dia}}(t)$ their time-dependent expansion coefficients. By inserting eq. \ref{eq:diabatic_expansion} into the time-dependent Schr\"odinger equation, one obtains a set of coupled differential equations for the expansion coefficients in the diabatic representation
\begin{equation}
    \text{i}\hbar\frac{\text{d}c^{\mathrm{dia}}_n}{\text{d}t} = \sum_m c^{\mathrm{dia}}_m(t) \left[ H^{\mathrm{Exc-FMO}}_{nm}(\mathbf{R}(t)) - \text{i}\hbar D_{nm}(\mathbf{R}(t)) \right],
\end{equation}
where $H^{\mathrm{Exc-FMO}}_{nm}$ is a matrix element of the complete excitonic Hamiltonian of the FMO-LC-TDDFTB method (cf. Fig. \ref{fig:excitonic_hamiltonian}). The nonadiabatic coupling is limited to the LE and CT states on the same fragments, as they are adiabatic states obtained from LC-TDA-DFTB calculations. The nonadiabatic coupling between all other quasi-diabatic states is zero.

Due to the fact that we employ the Ehrenfest method, the nuclei are propagated on the mean-potential of all electronic states. Thus, the excited state forces are approximated as the weighted linear combination of the gradients of the quasi-diabatic states
\begin{equation}
    F_R = -\sum_n |c_n^{\mathrm{dia}}|^2 \nabla_R E_n.
\end{equation}

In order to improve our Ehrenfest methodology with a decoherence correction, we have implemented the collapse to-a-block (TAB) decoherence correction of Esch and Levine\cite{esch_state-pairwise_2020,esch_decoherence-corrected_2020,esch_accurate_2021} in our FMO-LC-TDDFTB dynamics framework. The TAB method assumes the exponential decay of the coherences of the electronic density matrix, which is influenced by the difference of the gradients of the eigenstates of the excited Hamiltonian. After each nuclear time step, the system undergoes a stochastic collapse procedure, which yields new expansion coefficients for the Ehrenfest method that correspond to a specific block density matrix. To this end, the target density matrix $\rho^d$ is defined as
\begin{equation}
	\rho^d_{nm} = \begin{cases}
		\rho^c_{nn}(t) & n=m \\ \rho^c_{nm}(t) e^{-\Delta t/\tau_{nm}} & n \neq m
	\end{cases},
\end{equation}
where $\rho^c(t)$ is the coherent Ehrenfest density matrix
\begin{equation}
	\rho^c_{nm}(t) = (c^{\mathrm{dia}}_n(t))^* c^{\mathrm{dia}}_m(t)
\end{equation}
and $\tau_{nm}$ is the decoherence time between a pair of quasi-diabatic electronic states
\begin{equation}
	\tau_{nm}^{-2} = \sum_{R}\frac{\left[ F^{\mathrm{avg}}_{n,R} - F^{\mathrm{avg}}_{m,R} \right]^2}{\hbar^2 \alpha_R}.
\end{equation}
Here, $F^{\mathrm{avg}}_{m,R}$ is the average force over a nuclear time step for the quasi-diabatic state $m$ and the degree of freedom $R$, and $\alpha_R$ is an atom specific decoherence parameter. The TAB algorithm generates a set of density matrices $\mathbf{\rho}^{\mathrm{block}}_a$ and corresponding weights $P^{\mathrm{block}}_a$ that follow the sum condition
\begin{equation}
	\mathbf{\rho}^d = \sum_a P^{\mathrm{block}}_a \mathbf{\rho}^{\mathrm{block}}_a.
\end{equation}
Each density matrix is a pure superposition of a subset of the quasi-diabatic states. After each classical time step, the system undergoes a stochastic collapses into one of the block density matrices, yielding new expansion coefficients for the electronic propagation of the Ehrenfest method. The collapse procedure has been extensively described in various publications, and thus, we refer to the works of Esch and Levine for detailed derivation\cite{esch_state-pairwise_2020,esch_decoherence-corrected_2020,esch_accurate_2021}.

%\textcolor{red}{Results of the dynamics} 
\subsection{Strong Light-Matter Interactions}
In order to allow for the description of the interactions between microcavities and molecules, we have combined our FMO-LC-TDDTB approach with a generalized Tavis-Cummings\cite{tavis_exact_1968,tavis_approximate_1969} (TC) Hamiltonian to facilitate the description of strong light-matter interactions between large molecular aggregates and photonic modes of microcavities\cite{einsele_fmo-lc-tddftb_2024}. This approach includes both the polaritonic interaction between molecular aggregates and photon modes and the excitonic interactions between the molecules, which are usually neglected in common approaches. 
%In the present section, we only give a short overview of the theoretical framework of the methodology, a detailed overview is available in our recent paper\cite{einsele_fmo-lc-tddftb_2024}.

To describe the total excited state wavefunction of this approach, eq. \ref{eq:wavefunc_fmo} is expanded to include the eigenstates of the electromagnetic field, yielding a new quasi-diabatic basis, which contains locally excited, charge-transfer and photonic basis states:
\begin{align*}
    \left|\Psi\right\rangle=&\sum_{I}^N \sum_{m}^{N_{\mathrm{LE}}} c_I^m \left|\mathrm{LE}_I^m \right\rangle+ \sum_{I}^{N} \sum_{J \neq I}^{N} \sum_{m}^{N_{\mathrm{CT}}} c_{I\rightarrow J}^m \left| \mathrm{CT}_{I \to J}^{m}\right\rangle \\
    &+ \sum_{\kappa} c_{\kappa} \left| \mathrm{Ph}_{\kappa} \right\rangle \numberthis
\end{align*}
Here, the first excited photonic state of the $\kappa$-th mode of the electric field of a microcavity is denoted by $\left| \mathrm{Ph}_{\kappa} \right\rangle$. 

To obtain the polaritonic excited states of the complete system, the full excited Hamiltonian is constructed. Thus, the energies of the quasi-diabatic basis states and the couplings between them are required. The light-matter coupling between the LE states and the photonic cavity modes is given by
\begin{equation}
\label{eq:le_mode_coupling}
	\left\langle \mathrm{LE}_{I}^{m}|H|\mathrm{Ph}_{\kappa}\right\rangle=\hbar \sqrt{\frac{\hbar \omega_{ph,\kappa}}{2\epsilon_0 V}} \mu_{\mathrm{eg}}^{m(I)} \cdot \vec{\epsilon}_{\kappa},
\end{equation}
where V is the quantized volume of the electromagnetic field, $\vec{\epsilon}_{\kappa}$ is the polarization vector of the $\kappa$-th mode of the electromagnetic field with the frequency $\omega_{\mathrm{ph},\kappa}$ and $\mu_{\mathrm{eg}}^{m(I)}$ is the transition dipole moment of the $m$-th LE state of monomer $I$. As the coupling to the photonic basis states is dependent on the transition dipole moment of a molecular excited state, the coupling of the CT states to the cavity modes can be assumed to be approximately zero, because they generally posses a negligible transition dipole moment
\begin{equation}
	\left\langle \mathrm{CT}_{I \to J}^{m}|\mathrm{H}|\mathrm{Ph}_{\kappa}\right\rangle \approx 0.
\end{equation} 
The diagonalization of the total Hamiltonian yields the polaritonic excited states. The intensities of the polaritonic states can then be calculated according to
\begin{equation}
	f_i = \sum_{\kappa}\omega_i |c_\kappa|^2,
\end{equation}
where $\omega_i$ is the energy of the eigenstate and $c_\kappa$ is the coefficient of the photonic basis state of the $\kappa$-th cavity mode.

\section{Example applications}\label{sec:results}
In this section, we present a few exemplary applications with DIALECT to showcase the various features and the capabilities of the software package. At first, we show the simulation of the surface hopping dynamics of \textit{cis}-stilbene in the framework of LC-TDDFTB. This is followed by the calculation of the polariton dispersion of a naphthalene cluster consisting of 202 monomer fragments. Finally, we demonstrate the simulation of the exciton transfer in an anthracene chain, which consists of 30 fragment molecules, employing FMO-LC-TDDFTB.
\subsection{Surface Hopping Dynamics of \textit{cis}-stilbene}
The photoisomerization dynamics of \textit{cis}-stilbene has been extensively studied using different quantum chemical methods for the calculation of the electronic structure\cite{ioffe_photoisomerization_2013,williams_unmasking_2021,karashima_ultrafast_2023,jira_sensitivity_2024}. However, the product ratio of the photoisomerization between $trans$-stilbene, $cis$-stilbene and 4a,4b-dihydrophenanthrene (DHP)  remains a subject for ongoing debate\cite{williams_unmasking_2021,karashima_ultrafast_2023,jira_sensitivity_2024}. 

In this work, we investigate the dynamics of this system as a proof of principle to showcase the capability of DIALECT in simulating nonadiabatic molecular dynamics.

\begin{figure}[t!]
    \centering
    \includegraphics[width=\linewidth]{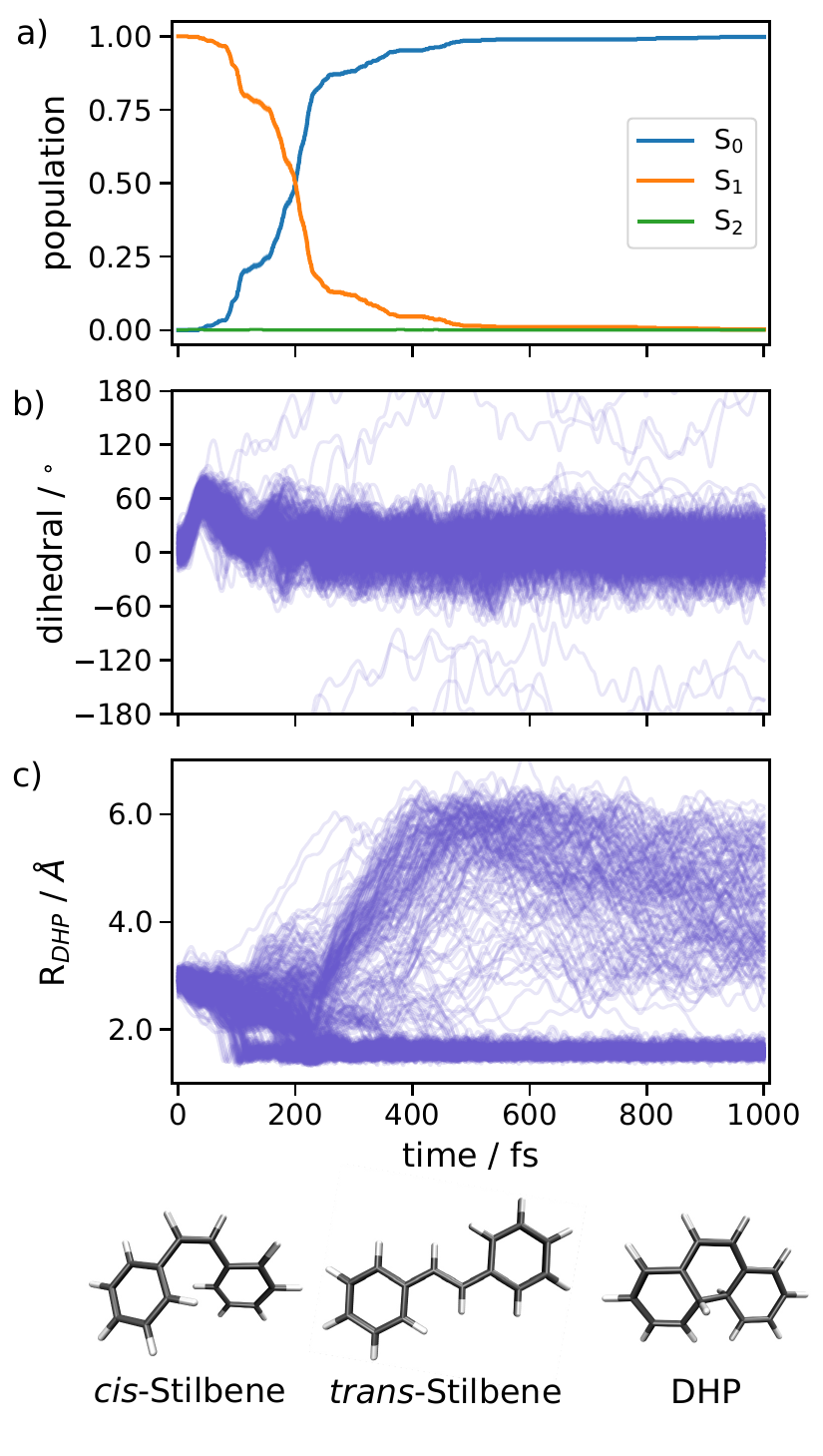}
    \caption{Nonadiabatic dynamics simulation of the photoisomerization of \textit{cis}-stilbene. a) Time evolution of the average electronic state populations. b) Time evolution of the dihedral angle between two phenyl groups for the vinylene bond for all trajectories. c) Time evolution of the 2-2' carbon-carbon distance ($R\mathrm{_{DHP}}$) for all trajectories.}
    \label{fig:stilbene-dynamics}
\end{figure}

The LC-DFTB2 method has been utilized in conjunction with the ob2\cite{vuong_parametrization_2018} parameter set for the calculation of the electronic structure. To generate the initial conditions for the simulation, a harmonic Wigner distribution was sampled at 200~K, yielding the initial positions and velocities for 500 trajectories. The nonadiabatic dynamics including four electronic singlet states (S$_0$, S$_1$, S$_2$ and S$_3$) were started in the bright S$_1$ state and were propagated for 1~ps with a timestep of 0.1~fs. The dynamics was performed by employing Tully's fewest-switch surface hopping procedure with the decoherence correction of Granucci et al\cite{granucci_critical_2007,granucci_including_2010}. The propagation of the ensemble of 500 trajectories takes on average 90~min for each trajectory on a single core of a Intel Xeon E5-2660 CPU, demonstrating the efficiency of our program.

The results of the nonadiabatic dynamics simulation are shown in Fig. \ref{fig:stilbene-dynamics}. After a small delay time, the population of the S$_1$ is rapidly transferred to the S$_0$ without involvement of the higher electronically excited states (cf. Fig \ref{fig:stilbene-dynamics}a). After approximately 540~fs, the population transfer from the S$_1$ to the electronic ground state is completed.

In order to track the structural changes of the trajectories and to differentiate between the three products of the photoisomerization reaction of \textit{cis}-stilbene, the time evolution of the dihedral angle of the vinylene bond and the 2-2' carbon-carbon distance are depicted in Fig. \ref{fig:stilbene-dynamics}b and \ref{fig:stilbene-dynamics}c for all 500 trajectories. 
After the start of the simulation, the vinylene dihedral shows a strong increase in the first 50~fs, while the R$_{DHP}$ distance remains relatively constant, which indicates the rapid twisting or pyramidalization of the vinylene double bond. After 100~fs, the trajectories start to diverge into two branches, the \textit{cis}-stilbene and DHP domains. This process is illustrated by the change in the $R\mathrm{_{DHP}}$ distribution for a large number of trajectories, which undergo a decrease of the $R\mathrm{_{DHP}}$ distance to ca. 1.6~\AA, which represents the formation of a C-C bond between the 2-2' positions of the phenyl rings. After 250~fs, the $R\mathrm{_{DHP}}$ distribution shows a strong bifurcation between trajectories which converge to the DHP domain, and trajectories that show a rapid increase in the $R\mathrm{_{DHP}}$ distance in conjunction with vinylene dihedral angles that strongly oscillate around zero degrees. The latter process indicates a strong motion of the phenyl rings of trajectories in the \textit{cis}-stilbene domain, which are caused by the transition from the S$_1$ to the S$_0$ state. From 400~fs onwards, the majority of the trajectories are moving on the ground state potential energy surface, while the last few trajectories finish the transition to the S$_0$ until 540~fs.
\begin{table}[t!]
\caption{Comparison of our LC-DFTB2 branching ratios of the photoisomerization of \textit{cis}-stilbene with published articles.}
\label{tab:branching_ratios}
\begin{tabular}{@{}lccc@{}}
\toprule
& \multicolumn{3}{c}{Branching ratio / \%}             \\ \midrule
\multicolumn{1}{c}{Method} & \textit{cis} & \textit{trans} &DHP \\ \midrule
LC-TDA-DFTB &   32.2   &   0.8    &    67       \\
OM3-MRCISD\cite{jira_sensitivity_2024} &    48       &      52           &     0           \\
XMS-SA3(2,2)-CASPT2\cite{karashima_ultrafast_2023} &    55       &      4         &     41         \\
XMS-SA3(2,2)-CASPT2\cite{jira_sensitivity_2024} &    55       &      2           &     43            \\
SA2(2,2)-CASSCF\cite{williams_unmasking_2021} &    45       &      52           &     4            \\
\bottomrule
\end{tabular}
\end{table}

The resulting branching ratio of the products of the simulated photoisomerization of \textit{cis}-stilbene is predicted to be strongly dominated by the \textit{cis}-stilbene and DHP domains, whereas only four out of 500 trajectories undergo an isomerization to the \textit{trans}-stilbene. A comparison of our results to the branching ratios of other works on the simulation of the stilbene photoisomerization is shown in Table \ref{tab:branching_ratios}.
While our prediction shows a qualitative agreement in regard to higher branching ratios of the \textit{cis}-stilbene and DHP structures in conjunction with a very low ratio for \textit{trans}-Stilben, as opposed to other semiempirical approaches, the results of our simulation with LC-DFTB still deviate from the branching ratios of other theoretical works that employ XMS-CASPT2. However, our semiempirical simulations agree with the predictions of the XMS-CASPT2 calculations that the ring closure of stilbene is favored over the isomerization of \textit{cis}-stilbene to \textit{trans}-stilbene. While the LC-DFTB2 method is only a SQM approach, and thus, cannot be compared to the quality of CASPT2 results, it can be utilized for the explorative investigation of photodynamics in molecular systems.

%\subsection{Excited-state spectrum of large systems}
\subsection{Polariton Dispersion}
To demonstrate the capability of DIALECT in simulating strong light-matter interactions between microcavities and large molecular aggregates, we investigate the polariton dispersion of a cluster of 203 naphthalene molecules (3654 atoms). To this end, a scan of the energy of the electric field cavity mode for different electric field polarization vectors is performed. The structure of the naphthalene aggregates has been generated from the crystal structure\cite{oddershede_charge_2004} with the Mercury program\cite{macrae_mercury_2020}. The ob2\cite{vuong_parametrization_2018} parameter set was employed for the DFTB calculations.

\begin{figure}[t!]
    \centering
    \includegraphics[width=\linewidth]{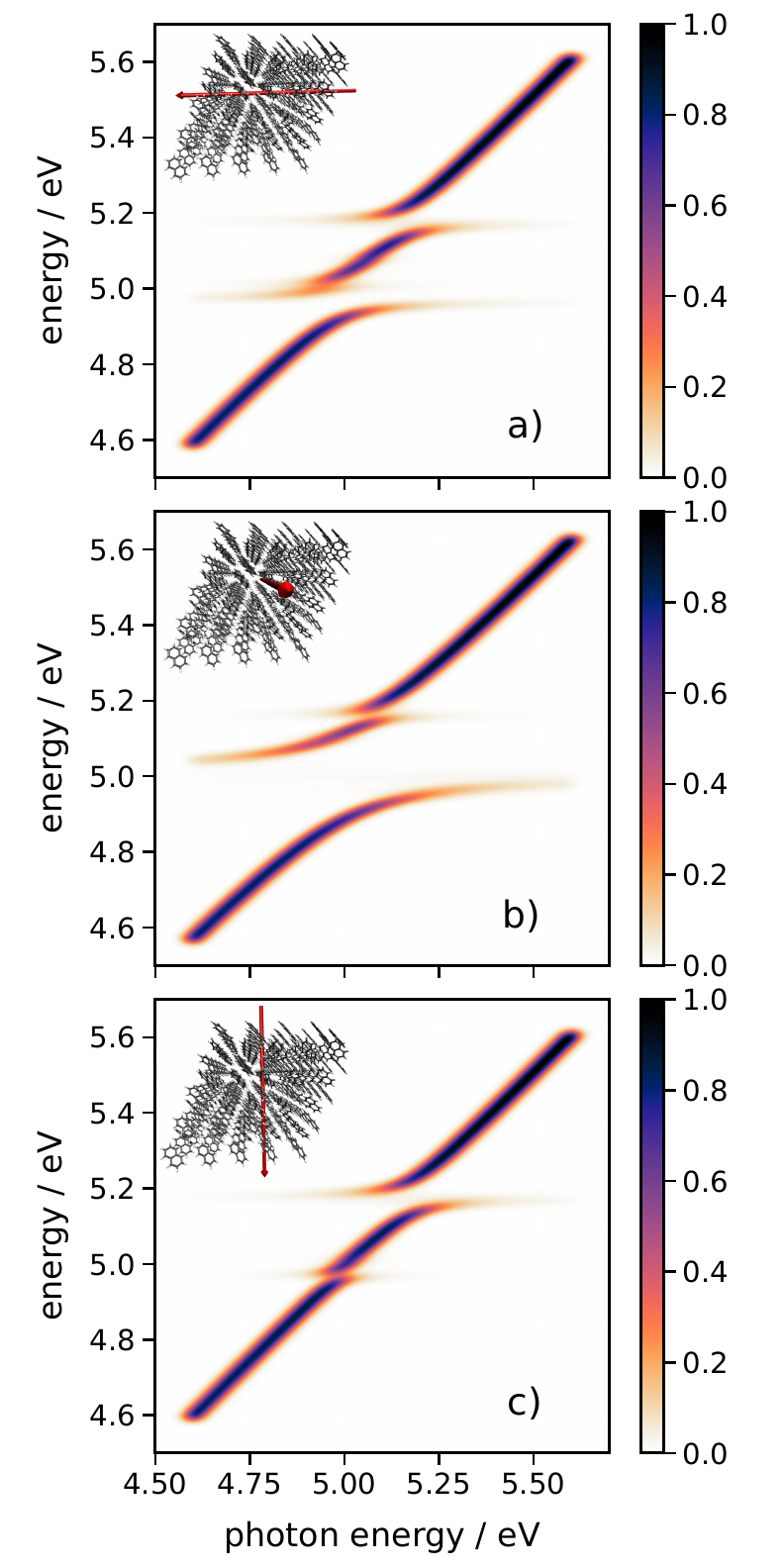}
    \caption{Polariton dispersion of an aggregate of naphthalene molecules. The energy of the cavity modes was scanned for a polarization of the electric field in a) x-direction, b) y-direction and c) z-direction.}
    \label{fig:polariton_dispersion}
\end{figure}
To simulate the polaritonic states, which stem from the coupling between the first few excited singlet states of the naphthalene aggregates and the cavity mode, the photon energy of the cavity is scanned between 4.6 and 5.6~eV. As the coupling strength of the light-matter interactions is strongly dependent on the length of the transition dipole moment components (cf. Eq. \ref{eq:le_mode_coupling}), the polariton dispersion is highly influenced by the polarization of the electric field vector. Due to the fact that on average the naphthalene molecules in the crystal possess the highest transition dipole component in the y-direction (cf. Tab. \ref{tab:transition_dipole_moments}), the polariton dispersion for a polarization of the electric field along the y-direction should have the highest coupling strength, leading to a strong splitting between the polaritonic branches.
\begin{table}[t!]
\caption{Absolute values of the transition dipole moment of a naphthalene monomer (in atomic units).}
\label{tab:transition_dipole_moments}
\begin{tabular}{c|ccc}
\toprule
state & x-axis & y-axis & z-axis  \\ \midrule
S$_1$  & 0.33  & 0.63  & 0.081 \\
S$_2$  & 0.19  & 0.072  & 0.26  \\ \bottomrule
\end{tabular}
\end{table}

The polariton dispersion of the naphthalene aggregates for polarizations in x, y and z-direction is depicted in Fig. \ref{fig:polariton_dispersion}. The polariton dispersion for the x-polarization shows two distinct splittings at photon energies around 5.0 and 5.1~eV, which can be attributed to the coupling between the first two excited singlet states of the naphthalene and the cavity mode. 

For the polarization along the y-axis, a large splitting between the upper and lower polaritonic branches can be observed for photon energies between 4.8 and 5.2~eV, which corresponds to the electronic transition of the naphthalene monomers to the S$_1$ state.
In addition, a small splitting occurs for a photon energy of ca. 5.1~eV, which results from the coupling between the S$_2$ states of the naphthalene fragments and the cavity mode. As the transition dipole moment of the S$_2$ along the y-axis is relatively low, the resulting light-matter coupling leads to a weak splitting between the lower and upper polaritonic branches.

The scan of the photon energy for the polarization in the z-direction shows the opposite picture. The coupling between the cavity mode and the S$_1$ states of the naphthalene aggregates is weak, while the coupling to the S$_2$ states results in a moderately strong splitting of the polaritonic. This distinction is also caused by the difference in the absolute value of the z-component of the transition dipole moment of the naphthalene molecules.

In summary, the simulation of the polariton dispersion of the naphthalene cluster shows the strong dependence of the light-matter coupling on the length of the transition dipole moment. By using the DIALECT program, one can investigate the coupling between large molecular assemblies and microcavities, including excitonic coupling effects between the individual molecules.
\subsection{Exciton Transfer Dynamics}
\begin{figure}[t!]
    \centering
    \includegraphics[width=\linewidth]{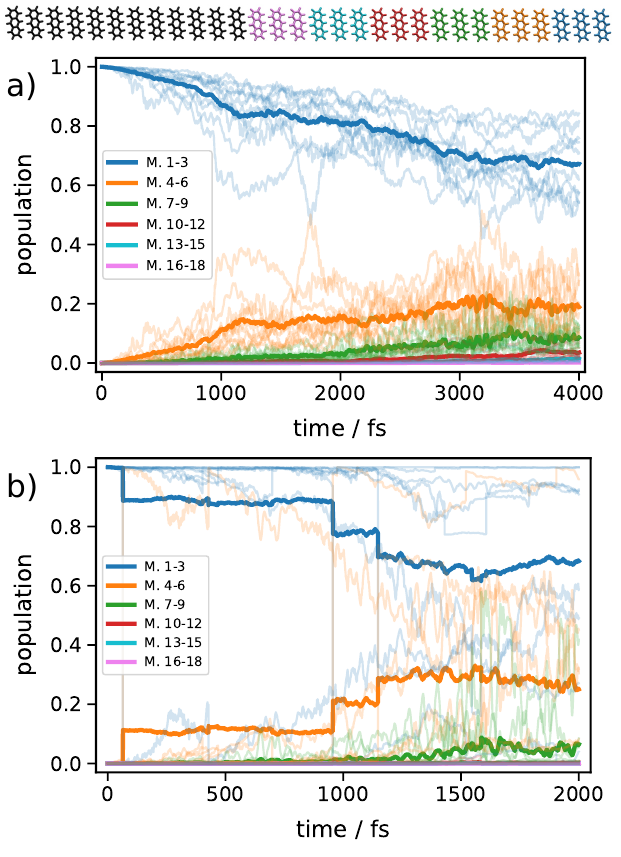}
    \caption{Averaged population contributions of the monomer fragments for the simulation of a) the standard Ehrenfest dynamics and b) the decoherence corrected Ehrenfest dynamics. The monomer contributions are separated in groups of three fragments for improved visibility. The populations of all trajectories are shown in the background.}
    \label{fig:anthracene_ehrenfest}
\end{figure}
To show the capabilities of DIALECT in simulating the exciton transfer dynamics in organic aggregates, the nonadiabatic excited state dynamics of an anthracene chain consisting of 30 molecules is investigated. To this end, we employ the FMO-LC-TDDFTB method in conjunction with the standard Ehrenfest method and the decoherence corrected Ehrenfest method\cite{esch_decoherence-corrected_2020}. In the framework of FMO-LC-TDDFTB, three LE states are used for each fragment monomer and one CT state for each fragment pair. The ob2 parameters set is used for the DFTB calculations and the anthracene monomer structure was generated from a crystal structure\cite{mason_crystallography_1964}. For the dynamics simulation, 10 trajectories are calculated for both Ehrenfest approaches, the timestep is set to 0.1~fs and the initial velocities of the trajectories were sampled from a Maxwell-Boltzmann distribution at 300~K. While the standard Ehrenfest trajectories were propagated for 40000 steps, the decoherence corrected Ehrenfest trajectories were propagated for 20000 steps.
The decoherence constant $\alpha$ was set to values of $40$~$a_0^{-2}$ for Hydrogen and $220$~$a_0^{-2}$ for Carbon.

The trajectories are initialized from the S$_1$ state of the outermost anthracene molecule. To prevent the anthracene molecules from drifting apart, we apply a weak harmonic restraint as an additional component to the nuclear forces.

We expect a gradual transfer of the initially localized exciton over the anthracene chain. To observe this process, the excited state population of the monomer fragments is averaged over all trajectories and depicted in Figs. \ref{fig:anthracene_ehrenfest}a and \ref{fig:anthracene_ehrenfest}b.  

\begin{figure*}[ht]
    \centering
    \includegraphics[width=\linewidth]{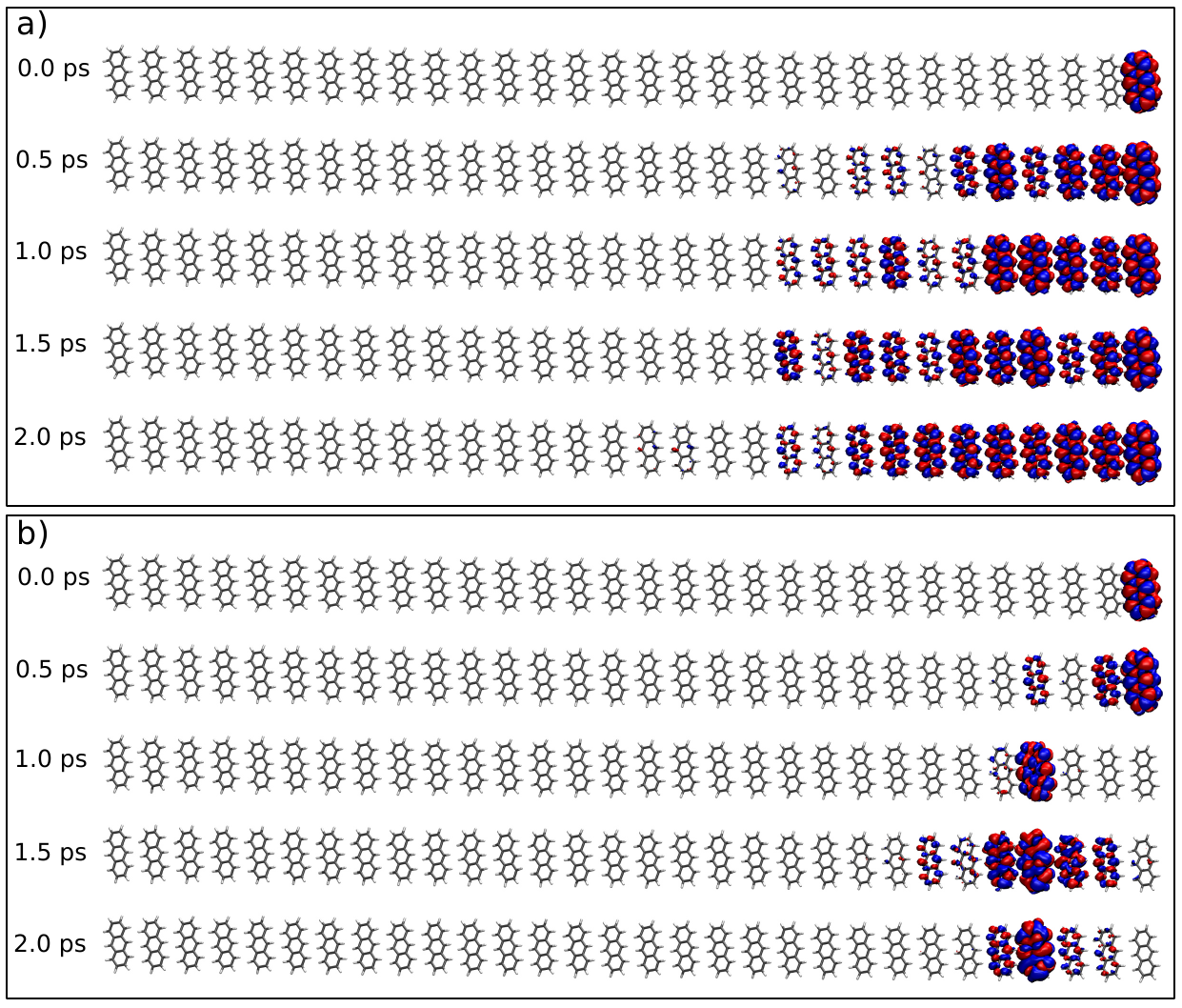}
    \caption{Transition densities of the excited state configuration at different points in time along two exemplary trajectories of a) the standard Ehrenfest approach and b) the decoherence corrected Ehrenfest approach.}
    \label{fig:anthracene_tdms}
\end{figure*}

The excited state population contributions of the monomers are split in groups of three molecules for improved visibility.
The simulations show a progressive transfer of the initial exciton to the neighboring anthracene fragments. The standard Ehrenfest (cf. Fig. \ref{fig:anthracene_ehrenfest}a) approach shows a slow delocalization of the excitation over the first 10 monomer fragments over a simulation time of 4000~fs, reaching very low contributions of the excited state population to even the 18th monomer. The trajectories of the decoherence corrected Ehrenfest dynamics show a much more localized transfer of the exciton over the anthracene chain. Along the different trajectories, the collapse into the various excited states can be observed in the time evolution of the excited monomer populations (cf. Fig. \ref{fig:anthracene_ehrenfest}b). The jumps in the population are caused by the collapse algorithm of the TAB procedure. 
The transition densities of the excited state configurations along two exemplary trajectories of both Ehrenfest approaches are depicted in Fig. \ref{fig:anthracene_tdms} to allow a visual comparison of the exciton transport. While the standard Ehrenfest dynamics can be characterized by the delocalized transport of the exciton over the anthracene chain (cf. Fig. \ref{fig:anthracene_tdms}a), the decoherence correction leads to a stronger localization of the exciton along the anthracene chain (cf. Fig. \ref{fig:anthracene_tdms}b). 

Thus, the decoherence correction of the Ehrenfest approach in the framework of the FMO-LC-TDDFTB method leads to the reduction of the strong delocalization of the standard Ehrenfest approach, which can overestimate the degree of delocalization in the transport of an exciton. To obtain a qualitative representation of the localized exciton dynamics for the anthracene chain, a larger number of trajectories is required. However, we present this example as a proof of concept for the implementation of the TAB procedure in the DIALECT program. We wish to point out that this procedure could be applied to calculate the exciton diffusion parameters from first principles, which will be a matter of further studies.
\section{Timings}\label{sec:timings}
This section will give a short summary of the performance of the DIALECT software package. To this end, we present the timings of DFTB and FMO-DFTB ground state calculations for water aggregates of various sizes. In addition, we show the performance of exemplary FMO-LC-TDDFTB calculations for clusters of tetracene molecules.

All benchmark calculations of this section were performed on a single node of a high performance computing cluster featuring two Intel Xeon E5-2680 CPUs and 180 Gb RAM.

The various crystal structures of the tetracene molecules were generated from a crystal structure\cite{holmes_nature_1999} by utilizing the Mercury software\cite{macrae_mercury_2020}.
\subsection{SCC Performance}
\begin{figure}
    \centering
    \includegraphics[width=\linewidth]{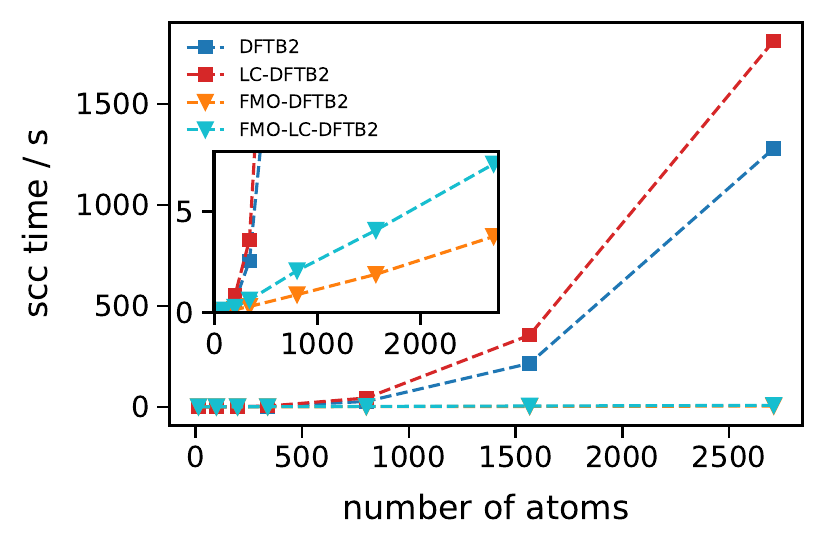}
    \caption{SCC timings for DFTB and FMO-DFTB calculations of various water clusters, containing 5, 33, 66, 113, 267, 522 and 903 $\mathrm{H_2O}$ molecules (15, 99, 198, 339, 801, 1566 and 2709 atoms). The inset shows the SCC times up to 8 seconds to differentiate between FMO-DFTB2 and FMO-LC-DFTB2.}
    \label{fig:water_scc_timings}
\end{figure}

In order to show the performance of the DIALECT software with the FMO-DFTB method and standard DFTB calculations, ground state SCC timings of various water clusters of different sizes (5, 33, 66, 113, 267, 522 and 903 $\mathrm{H_2O}$ molecules) are investigated. In addition, the differences in computational demand between the DFTB and LC-DFTB approaches are shown. The results of this comparison are depicted in Fig. \ref{fig:water_scc_timings}. 
In comparison to the standard DFTB method, the near linear scaling of the FMO-(LC)-DFTB method leads to substantially lower computational resources, requiring 0.3\% of the wall time of the DFTB method for calculation of the water cluster consisting of 903 molecules.

\subsection{Excited State Performance}
\begin{figure}
    \centering
    \includegraphics[width=\linewidth]{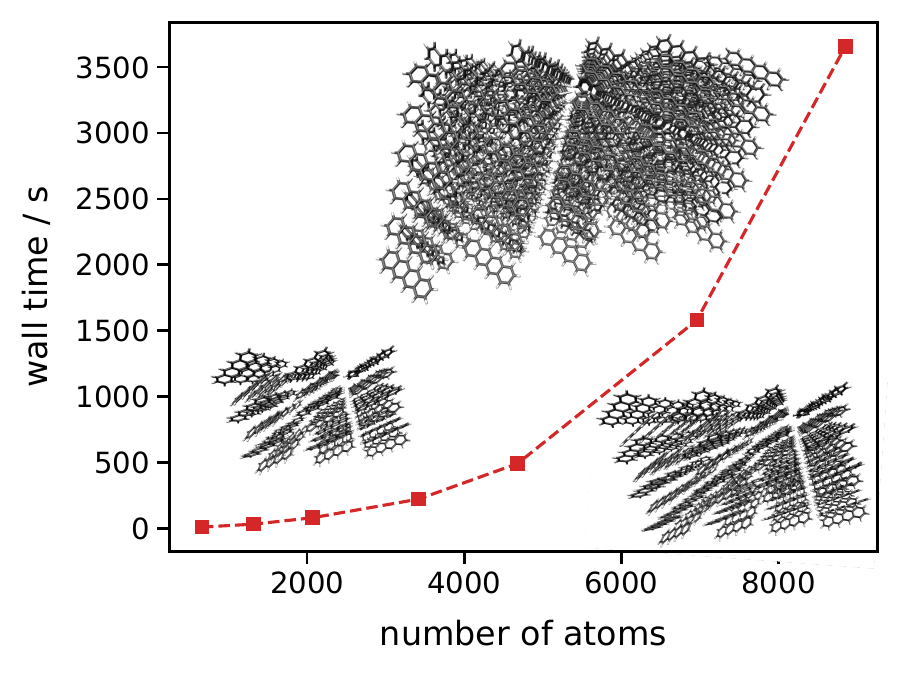}
    \caption{Computation timings of FMO-LC-TDDFTB calculations of seven tetracene aggregates, consisting of up to 8850 atoms.}
    \label{fig:tetracene_benchmark}
\end{figure}

The performance of the DIALECT program for the computation of the excited states of large molecular aggregates is investigated by exemplary calculations of various tetracene assemblies, containing up to 295 monomers. To this end, the required calculation times for seven different tetracene structures (660, 1320, 2070, 3420, 4680, 6960 and 8850 atoms) were investigated. The FMO-LC-TDDFTB calculations involved four locally excited states for each fragment and one charge-transfer state for each fragment pair. The calculation of the excited state manifold of the complete system was limited to the first 200 energetically lowest states for each tetracene assembly. The timings of the FMO-LC-TDDFTB calculations are depicted in Fig. \ref{fig:tetracene_benchmark}.
As already shown in a previous work, the excited state calculation of large molecular assemblies does not have the same linear scaling behavior of the ground-state calculation. The calculation of the excitonic couplings and the subsequent diagonalization of the excitonic Hamiltonian by the Davidson method lead to worse computational scaling.\cite{einsele_long-range_2023}

However, by using the DIALECT software, the calculation of molecular aggregates, which consist of over a 10000 atoms, is possible in a fast and efficient manner. As shown in Fig. \ref{fig:tetracene_benchmark}, a calculation of the excited states of 295 tetracene molecules requires only approximately one hour. Thus, by using the DIALECT software, the study of the excited state properties of organic semiconductors or large biomolecular aggregates is feasible.  
\section{Conclusion and Outlook}\label{sec:conclusion}
With DIALECT, we have developed a program package that is capable of simulating the ground and excited state properties of large excitonic systems, including nonadiabatic molecular dynamics and light-matter interactions between microcavities and molecular assemblies. By utilizing the DIALECT software, the investigation of the exciton and charge transfer dynamics in large molecular assemblies is feasible due to the fast and efficient (LC-)DFTB and FMO-(LC-)DFTB framework implemented in our program package.

As demonstrated in this work, DIALECT can be utilized to calculate nonadiabatic molecular dynamics both with the (LC-)TDDFTB and FMO-(LC-)TDDFTB methods. The exemplary nonadiabatic dynamics of \textit{cis}-stilbene shows that DIALECT can provide a very fast and efficient investigation of the photodynamics of molecular systems.
Although a high accuracy cannot be expected from a DFTB based method,  due to the efficiency insights into the excited state dynamics can be obtained at a low price.
By combining the TAB decoherence correction with the Ehrenfest approach in the FMO-LC-TDDFTB framework, the simulation of localized exciton transport in large molecular assemblies becomes possible.

In addition, with DIALECT, the computation of the excited state spectra of molecular aggregates, which are comprised of over 10000 atoms becomes feasible. Furthermore, DIALECT can be used to simulate strong light-matter interactions between microcavities and molecular assemblies, including the excitonic couplings between the molecular fragments.

In the future, we plan to investigate the exciton dynamics in large molecular aggregates, which are influenced by the strong light-matter coupling to cavities. We are also working on the release of a version of DIALECT that is based on ab-initio quantum mechanical methods to improve upon the semiempirical treatment inherent in DFTB. 
Ultimately, our program should enable first-principles atomistic simulations of energy and charge-transfer dynamics in large biomolecular systems or realistic models for optoelectronic devices.

%%%%%%%%%%%%%%%%%%%%%%%%%%%%%%%%%%%%%%%%%%%%%%%%%%%%%%%%%%%%%%%%%%%%%
%% The "Acknowledgement" section can be given in all manuscript
%% classes.  This should be given within the "acknowledgement"
%% environment, which will make the correct section or running title.
%%%%%%%%%%%%%%%%%%%%%%%%%%%%%%%%%%%%%%%%%%%%%%%%%%%%%%%%%%%%%%%%%%%%%
\begin{acknowledgement}
We gratefully acknowledge financial support by the Deutsche Forschungsgemeinschaft (DFG, German Research Foundation) via the grants MI1236/7-1 and IRTG 2991 Photoluminescence in Supramolecular Matrices – Project number 517122340. L.N.P acknowledges the support by the FCI and X.M. thanks the Studienstiftung des deutschen Volkes for the fellowship.
\end{acknowledgement}

%%%%%%%%%%%%%%%%%%%%%%%%%%%%%%%%%%%%%%%%%%%%%%%%%%%%%%%%%%%%%%%%%%%%%
%% The same is true for Supporting Information, which should use the
%% suppinfo environment.
%%%%%%%%%%%%%%%%%%%%%%%%%%%%%%%%%%%%%%%%%%%%%%%%%%%%%%%%%%%%%%%%%%%%%
\begin{suppinfo}
The data that supports this work is available at \url{https://github.com/mitric-lab/Data_DIALECT_software_package_paper}. The structures of the molecular assemblies for the benchmark calculations and the polariton mode frequency scan data are provided.  

% A listing of the contents of each file supplied as Supporting Information
% should be included. For instructions on what should be included in the
% Supporting Information as well as how to prepare this material for
% publications, refer to the journal's Instructions for Authors.
% The following files are available free of charge.
% \begin{itemize}
%   \item Filename: brief description
%   \item Filename: brief description
% \end{itemize}

\end{suppinfo}

%\newpage
%%%%%%%%%%%%%%%%%%%%%%%%%%%%%%%%%%%%%%%%%%%%%%%%%%%%%%%%%%%%%%%%%%%%%
%% The appropriate \bibliography command should be placed here.
%% Notice that the class file automatically sets \bibliographystyle
%% and also names the section correctly.
%%%%%%%%%%%%%%%%%%%%%%%%%%%%%%%%%%%%%%%%%%%%%%%%%%%%%%%%%%%%%%%%%%%%%
\bibliography{references}

\end{document}